\documentclass[showpacs,aps,twocolumn]{revtex4} 
\usepackage{graphicx,amsmath,dcolumn}

\newcommand{\ud}{\mathrm{d}}
\newcommand{\jn}{\mathcal{J}_{N}}
\newcommand{\ju}{\mathcal{J}_{U}}
\newcommand{\force}{\mathcal{A}}
\newcommand{\dv}{\Delta}
\begin{document} 
\title{Fluctuation theorem for the effusion of an ideal gas.} 
\author{B. Cleuren}
\email{bart.cleuren@uhasselt.be}
\affiliation{Hasselt University - B-3590 Diepenbeek, Belgium}
\author{C. \surname{Van den Broeck}}
\affiliation{Hasselt University - B-3590 Diepenbeek, Belgium}
\author{R. Kawai}
\affiliation{Department of Physics, University of Alabama at Birmingham,
Birmingham, Alabama 35294, USA} 

\begin{abstract}
The probability distribution of the entropy production for the effusion of an ideal gas between two compartments is calculated explicitly. The fluctuation theorem is verified. The analytic results are in good agreement with numerical data from hard disk molecular dynamics simulations.
\end{abstract}
\date{\today}
\maketitle

\section{Introduction}
Effusion is the motion of a gas through a small pore or opening. The linear dimension of the opening is assumed to be smaller than the mean free path of the gas particles. Consequently, the leak does not disturb the state of the gas. If the latter is at equilibrium, it will remain so and many properties of the effusing gas particles can be easily calculated, using basic arguments from kinetic theory of gases. One of the older results is the so-called Graham's law \cite{graham}, stating that the rate of effusion is inversely proportional to the square root of the mass of the gas particles. This observation was in fact used in the production of the atomic bom during the Manhattan project for the enrichment of uranium. Graham's law is a direct consequence of the obvious fact that faster particles will exit through the hole more frequently. In fact, the average kinetic energy of the escaping particles turns out to be $2kT$, to be compared to the bulk average value of $(3/2)kT$ (3 dimensional system). As a result, effusion will lead to a cooling of the gas, a principle that is used to date for refrigeration \cite{refrigeration}. In the early 1900's, Knudsen investigated other aspects of effusion \cite{knudsen}. He discovered the cosine law, corresponding to the fact that the particles exit with an isotropic angular distribution, a property which was subsequently using for coating spherical bulbs, and which is also related to the Kirchoff's laws of radiation and to proper boundary condition for particles that are thermally reemitted from an absorbing wall \cite{garciabook}.

\begin{figure}[t]
\begin{center}
\includegraphics[width=0.3\textwidth]{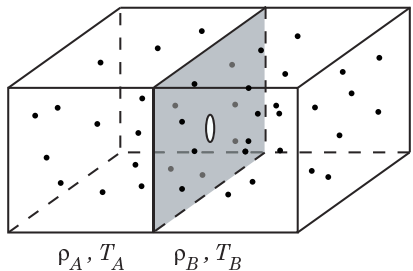}
\caption{The system under consideration: two ideal gases, in equilibrium at their respective temperatures and densities, can exchange energy and particles through a small hole in the common adiabatic wall.}
\label{figSetup}
\end{center}
\end{figure}

An interesting situation occurs when the small hole forms the contact between two containers (referred to as Knudsen cells) with gases which are separately at equilibrium, but not at equilibrium with each other, cf. Fig. ~\ref{figSetup}. While  local equilibrium (inside each container) is preserved, the existing density and temperature gradients, which can be arbitrarily large, will lead to both particle and energy fluxes. A somewhat surprising observation, already noted by Knudsen, is that for $\rho_A \sqrt{T_A}=\rho_B \sqrt{T_B}$ an energy flux is present in the absence of a corresponding particle flux, cf. Eq.~(\ref{fluxesParticles}) (this peculiar property also appears in the so-called Jepsen gas \cite{bala}). Similarly, a  particle flux is present with no corresponding energy flux when $\rho_A {T_A}^{3/2}=\rho_B {T_B}^{3/2}$, cf. Eq.~(\ref{fluxesEnergy}). When operating in the regime of weak gradients, more precisely in the framework of linear irreversible thermodynammics, the properly defined  thermodynamic forces and fluxes are related to each other by a symmetric Onsager matrix \cite{onsager,prigogine,degroot}.

The above properties refer to average values of the fluxes. It is however clear that one can easily calculate fluctuations and correlation functions. For example, the typical appearance of long range spatial correlations in nonequilibrium systems was illustrated in detail for an array of Knudsen cells \cite{baras,malek,garcia,nicolis}.

In the present paper, we go one step further and calculate the full joint probability distribution for the  particle and energy flux. This result is particularly relevant because of the recent discovery of various work and fluctuation theorems \cite{bochkov,evans1,gallavotti,jarzynski,kurchan,lebowitz,crooks,maes,evansREVIEW,evans2,derrida2,seifertPRL2005}, which result from a time reversal symmetry of the underlying dynamics. In the present case, it implies that
the probability density  $P_{t}(\dv S)$ for the entropy $\dv S$, produced by effusion during a time interval of duration $t$, satisfies the following relation:
\begin{equation}\label{ft}
\frac{P_{t}(\dv S)}{P_{t}(-\dv S)}=e^{\dv S/k}.
\end{equation}
This result implies:
\begin{equation}\label{crooks}
\langle e^{-\dv S/k}\rangle=1,
\end{equation}
and consequently (invoking Jensen's inequality) $\langle \dv S \rangle \geq 0$, in agreement with the second law.

The purpose of this paper is to present an explicit and  detailed analysis of effusion in the context of various fluctuation theorems. Their verification is most easily done at the level of the cumulant generating function. The effusion problem moreover has the pleasant feature that the marginal distribution functions for particle and energy transport can also be calculated explicitly. They obey separate fluctuation theorems when the thermodynamic force for the other transport process vanishes. Their  large deviation functions can be calculated explicitly and are related by Legendre transformation to the cumulant generating function.
Finally, we report  the explicit results for the two-dimensional version,  allowing a detailed comparison with numerical data from hard disk molecular dynamics simulations.

\section{Fluctuation theorem for effusion}
Consider two large reservoirs, $A$ and $B$, separated by a common adiabatic wall, each containing an ideal gas at equilibrium, with  uniform density $\rho_{i}$ and  Maxwellian velocity distribution $\phi_{i}(\vec{v})$ at temperature $T_{i}$, $i \in\{A,B\}$  (cf. Fig.~\ref{figSetup}):
\begin{equation}
\phi_{i}(\vec{v})=\left(\frac{m}{2\pi kT_{i}}\right)^{\frac{3}{2}}e^{-\frac{mv^{2}}{2kT_{i}}}.
\end{equation}
At time $t=0$ , a small hole  with surface area $\sigma$ is opened, allowing an exchange during a time interval of length $t$, of particles and energy between the two compartments. The dimensions of the hole are assumed small compared to the mean free path of the gas particles, and the reservoirs sufficiently large so that the equilibrium state in both parts is not affected by the exchange. In particular, the temperatures and densities in both reservoirs remain constant \footnote{The case of finite reservoirs, in which density and energy follows adiabatically the exchange through the hole, can also be solved, but is not directly relevant for the present discussion.}. Hence, the change in entropy in the total system,
upon a total transfer of energy $\dv U$ and of particles $\dv N$ from $A$ to $B$ during $t$, is given by 
\begin{eqnarray}
\dv S&=& \dv S_{A}+ \dv S_{B}
\nonumber \\&=&-\frac{1}{T_{A}}\dv U+\frac{\mu_{A}}{T_{A}}\dv N+\frac{1}{T_{B}}\dv U-\frac{\mu_{B}}{T_{B}}\dv N \nonumber \\&=&\force_{U}\dv U +\force_{N}\dv N.
\label{eq:entropy}
\end{eqnarray}
We introduced, in accordance with the definitions from irreversible thermodynamics, the following thermodynamic forces (affinities) for energy and particle flow, respectively:
\begin{eqnarray}
\force_{U}&=&\frac{1}{T_{B}}-\frac{1}{T_{A}};\nonumber \\
\force_{N}&=&\frac{\mu_{A}}{T_{A}}-\frac{\mu_{B}}{T_{B}}=k\log\left[\frac{\rho_{A}}{\rho_{B}}\left(\frac{T_{B}}{T_{A}}\right)^{\frac{3}{2}}\right].
\label{eq:tforces}
\end{eqnarray}
In the last equality, we used the  expression for the chemical potential of an ideal gas. Note that the thermodynamic force $\force_{N}$ diverges when one of the (infinitely large!) reservoirs is empty, so that free effusion into unlimited space, implying an infinitely large entropy production, corresponds to a singular limit.

Clearly, as single particle crossings from both sides contribute to the values of  $\dv U$ and $\dv N$ in the course of time, these quantities  are fluctuating, and hence so is the entropy production $\dv S$. As indicated in the introduction, the resulting probability density for the  entropy production should obey the fluctuation theorem, Eq.~(\ref{ft}). Clearly, the contributions to $\dv S$ from any two equal but  non-overlapping time intervals are independent and identically distributed random variables. In other words, $\dv S$ is a stochastic process with independent increments. As a result its cumulant generating function has the following form:
\begin{equation}
\langle e^{-\lambda \dv S}\rangle=e^{-t \mu(\lambda)}.
\end{equation}
The fluctuation theorem Eq.~(\ref{ft}) implies the following symmetry relation for the function $\mu(\lambda)$ (to which we will also refer, for brevity, as the cumulant generating function):
\begin{equation}
\mu(\lambda)=\mu(k^{-1}-\lambda).
\label{FT}
\end{equation}
We will in fact prove for the system under consideration a more detailed fluctuation theorem, expressed in terms of the joint probability density, namely:
\begin{equation}\label{dft}
\frac{P_{t}(\dv U,\dv N)}{P_{t}(-\dv U,-\dv N)}=e^{\dv S/k}.
\end{equation}
The increments of $\dv U$ and $\dv N$ are again independent so that the corresponding cumulant generating function has the form:
\begin{equation}
\langle e^{-(\lambda_{U}\dv U+\lambda_{N}\dv N)}\rangle=e^{-t\mu(\lambda_{U},\lambda_{N})}.
\end{equation}
In terms of the function $\mu(\lambda_{U},\lambda_{N})$, the detailed fluctuation theorem reads:
\begin{equation}
\mu(\lambda_{U},\lambda_{N})=\mu(\force_{U}/k-\lambda_{U},\force_{N}/k-\lambda_{N}).
\label{DFT}
\end{equation}
Since $\dv S=\force_{U}\dv U +\force_{N}\dv N$, one has $\mu(\lambda)=\mu(\lambda \force_{U},\lambda \force_{N})$, so that the detailed fluctuation theorem implies the "normal" fluctuation theorem. 

Eq.~(\ref{dft}) also implies fluctuation theorems for energy and particle transport  separately, when the corresponding thermodynamic force for the other process vanishes. Indeed,  in this case only  one of the variables,  $\dv U$ or $\dv N$,  appears in $\dv S$. Bringing the denominator from lhs to rhs  in Eq.~(\ref{dft}), subsequent integration over the other variable implies:
\begin{equation}\label{dft1}
\frac{\mathcal{P}_{t}(\dv U)}{\mathcal{P}_{t}(-\dv U)}=e^{\dv S/k} \;\;\;\;\;\; \mbox{when} \;\;\;\;\;\; \force_N=0,
\end{equation}
and 
\begin{equation}\label{dft2}
\frac{\mathcal{P}_{t}(\dv N)}{\mathcal{P}_{t}(-\dv N)}=e^{\dv S/k} \;\;\;\;\;\; \mbox{when} \;\;\;\;\;\; \force_U=0.
\end{equation}
We now proceed to a direct verification of the detailed fluctuation theorem Eq.~(\ref{DFT}) and its implications Eqs. (\ref{FT},\ref{dft1},\ref{dft2}) by an explicit evaluation of the generating function $\mu(\lambda_{U},\lambda_{N})$.

\section{Master equation and cumulant generating function}
During a small time interval $dt$, the contributions to quantities $\dv U$ and $\dv N$ result from individual particle transport across the hole. Following basic arguments from kinetic theory of gases (for more details, see Appendix \ref{transition}), the corresponding probabilities per unit time, $T_{A \rightarrow B}(E)$ and $T_{B \rightarrow A}(E)$, to observe a particle with kinetic energy $\frac{1}{2}mv^{2}=E$ crossing the hole from $A \rightarrow B$ and $B \rightarrow A$ respectively, are given by:
\begin{eqnarray}\label{eq:tr}
T_{A \rightarrow B}(E)&=&\frac{\sigma \rho_{A}}{\sqrt{2\pi m k T_{A}}}\frac{E}{kT_{A}}e^{-\frac{E}{kT_{A}}} ; \nonumber \\
T_{B \rightarrow A}(E)&=&\frac{\sigma \rho_{B}}{\sqrt{2\pi m k T_{B}}}\frac{E}{kT_{B}}e^{-\frac{E}{kT_{B}}}.
\end{eqnarray}
The probability density $P_{t}(\dv U,\dv N)$ thus obeys the following Master equation:
\begin{multline}
\frac{\partial}{\partial t}P_{t}(\dv U,\dv N)=\int_{0}^{\infty} T_{A \rightarrow B}(E)P_{t}(\dv U-E,\dv N-1)\ud E \\
+\int_{0}^{\infty} T_{B \rightarrow A}(E)P_{t}(\dv U+E,\dv N+1)\ud E  \\
-P_{t}(\dv U,\dv N)\int_{0}^{\infty}\left[T_{A \rightarrow B}(E)+T_{B \rightarrow A}(E)\right]\ud E.
\label{eq:master}
\end{multline}
Note that the integral operators are of the convolution type, in agreement with the fact that the processes $\dv U$ and $\dv N$ have independent increments.
Hence, the exact solution of Eq.~(\ref{eq:master}), subject to the initial condition $P_{0}(\dv U,\dv N)=\delta (\dv U)\delta_{\dv N,0}$, is obtained by Fourier transfom, i.e., by switching to the cumulant generating function $\mu(\lambda_{U},\lambda_{N})$. Since the following integrals can be performed explicitly:
\begin{align}
\int_{0}^{\infty} T_{A \rightarrow B}(E)e^{-\lambda_{U}E}\ud E&=
\frac{\sigma}{\sqrt{2\pi m}}\frac{\rho_{A}\sqrt{kT_{A}}}{(1+k T_{A}\lambda_{U})^{2}};\nonumber \\
\int_{0}^{\infty} T_{B \rightarrow A}(E)e^{\lambda_{U}E}\ud E&=
\frac{\sigma}{\sqrt{2\pi m}}\frac{\rho_{B}\sqrt{kT_{B}}}{(1-k T_{B}\lambda_{U})^{2}},
\end{align}
we immediately obtain:
\begin{multline}
\mu(\lambda_{U},\lambda_{N})=\frac{\sigma \sqrt{k}}{\sqrt{2\pi m}}\left(\rho_{A}\sqrt{T_{A}}\left[1-\frac{e^{-\lambda_{N}}}{(1+kT_{A}\lambda_{U})^{2}}\right] \right. \\ \left. +\rho_{B}\sqrt{T_{B}}\left[1-\frac{e^{\lambda_{N}}}{(1-kT_{B}\lambda_{U})^{2}}\right]
\right).
\label{eq:mu}
\end{multline}
Eq.~(\ref{eq:mu}) is a central result of this paper. One easily verifies, using the explicit expressions for the thermodynamic forces given in Eq.~(\ref{eq:tforces}), that this expression indeed verifies the detailed fluctuation theorem Eq.~(\ref{DFT}). 

Note also that $\mu(\lambda_{U},\lambda_{N})$ can be written as the sum of two contributions:
\begin{equation}\label{eq:mua}
\mu(\lambda_{U},\lambda_{N})=\mu_{A}(\lambda_{U},\lambda_{N})+\mu_{B}(-\lambda_{U},-\lambda_{N}),
\end{equation}
with 
\begin{eqnarray}\label{mou}
\mu_{A}(\lambda_{U},\lambda_{N})&=&\frac{\sigma \rho_{A}\sqrt{kT_{A}}}{\sqrt{2\pi m}}\left[1-\frac{e^{-\lambda_{N}}}{(1+kT_{A}\lambda_{U})^{2}}\right],\;\;\;
\end{eqnarray}
and the similar expression for A changed into B. This additivity property derives from the statistical independence of the fluxes from $A \rightarrow B$ and $B \rightarrow A$.

\section{Cumulants}
The joined cumulant  $\kappa_{ij}$ of power $i$ in energy flux and $j$ in particle flux appears as a coefficient of the Taylor expansion of the cumulant generating function, namely:
\begin{equation}
\mu(\lambda_{U},\lambda_{N})=-\frac{1}{t}\sum_{i,j=0}^{\infty}\frac{(-1)^{i+j}\lambda_{U}^{i}\lambda_{N}^{j}}{i!j!}\kappa_{ij}.
\end{equation}
The explicit expressions for the cumulants of energy and particle flux can thus easily be obtained from Eq.~(\ref{eq:mu}).
We mention the following results. The first order cumulants read:
\begin{eqnarray}
\kappa_{10}&=&\langle \dv U \rangle= \frac{t \sigma k^{3/2}}{\sqrt{2\pi m}}2\left(\rho_{A}T_{A}^{3/2}-\rho_{B}T_{B}^{3/2} \right); \label{cum1}\\
\kappa_{01}&=&\langle \dv N \rangle= \frac{t \sigma k^{1/2}}{\sqrt{2\pi m}}\left(\rho_{A}T_{A}^{1/2}-\rho_{B}T_{B}^{1/2} \right).\label{cum2}
\end{eqnarray}
The second order  cumulants are:
\begin{eqnarray}
\kappa_{20}&=&\langle \delta \dv U^{2} \rangle= \frac{t \sigma k^{5/2}}{\sqrt{2\pi m}}6\left(\rho_{A}T_{A}^{5/2}+\rho_{B}T_{B}^{5/2} \right);\\
\kappa_{11}&=& \frac{t \sigma k^{3/2}}{\sqrt{2\pi m}}2\left(\rho_{A}T_{A}^{3/2}+\rho_{B}T_{B}^{3/2} \right);\\
\kappa_{02}&=&\langle \delta \dv N^{2} \rangle= \frac{t \sigma k^{1/2}}{\sqrt{2\pi m}}\left(\rho_{A}T_{A}^{1/2}+\rho_{B}T_{B}^{1/2} \right).
\end{eqnarray}
Similar expressions can be obtained for the higher order cumulants. Note that all cumulants are proportional to time, the specifying feature of processes with independent increments.

\section{Marginal distributions}
The explicit calculation of $P_{t}(\dv U,\dv N)$ is quite involved, but  it is relatively easy to obtain analytic expressions for the marginal distributions:
\begin{eqnarray}
\mathcal{P}_{t}(\dv N)&=&\int_{-\infty}^{+\infty}P_{t}(\dv U,\dv N)\ud \dv U; \\
\mathcal{P}_{t}(\dv U)&=&\sum_{\dv N=-\infty}^{+\infty}P_{t}(\dv U,\dv N).
\end{eqnarray}
The corresponding cumulant generating functions of these marginal distributions are obtained by setting $\lambda_{N}=0$ and $\lambda_{U}=0$ respectively in Eq.~(\ref{eq:mu}).

The generating function for particle transport is thus found te be:
\begin{multline}\label{mon}
\mu(0,\lambda)=\frac{\sigma \sqrt{k}}{\sqrt{2\pi m}}\left(\rho_{A}\sqrt{T_{A}}\left[1-e^{-\lambda}\right]\right. \\ + \left.\rho_{B}\sqrt{T_{B}}\left[1-e^{\lambda}\right]
\right).
\end{multline}
This result is identical to that for a  random walk taking steps to the right and left with probabilities per unit time:
\begin{align}
r_{A}&=\sigma \rho_{A}\sqrt{kT_{A}/2\pi m};\\
r_{B}&=\sigma \rho_{B}\sqrt{kT_{B}/2\pi m},
\end{align}
respectively. This relation is intuitively clear since the passages of  particles from reservoir $A$ to $B$ and vice-versa induce a random walk on the variable $\dv N$. A similar expression for the cumulant generating function, but with  $r_{B}=0$, appears in the problem of shot noise \cite{derrida1, blanter}. The probability distribution $\mathcal{P}_{t}(\dv N)$ is, as is well known from the random walk literature, expressed in terms of the modified Bessel function. Since we have:
\begin{equation}
e^{-\tau \mu(0,\lambda)}=\sum_{\Delta N}e^{-\lambda \Delta N}\mathcal{P}_{t}(\dv N),
\end{equation}
and using the generating function of the modified Bessel functions \cite{abramowitz}:
\begin{equation}
e^{\frac{1}{2}z(e^{-\lambda}+e^{\lambda})}=\sum_{n}e^{-\lambda n}I_{n}(z),
\end{equation}
one finds:
\begin{widetext}
\begin{equation}\label{bessel}
\mathcal{P}_{t}(\dv N)=e^{-\frac{t\sigma }{\sqrt{2\pi m}}\left(\rho_{A}\sqrt{kT_{A}}+\rho_{B}\sqrt{kT_{B}}
\right)}\left(\frac{\rho_{A}\sqrt{T_{A}}}{\rho_{B}\sqrt{T_{B}}}\right)^{\dv N/2} I_{\dv N}\left(t\sigma \sqrt{\frac{2k}{\pi m}}(\rho_{A}\rho_{B}\sqrt{T_{A}T_{B}})^{1/2}\right).
\end{equation}
\end{widetext}
When $T_{A}=T_{B}$, i.e., when the thermodynamic force for the energy flow vanishes, $\force_{U}=0$, we can verify explicitly  the fluctuation theorem Eq.~(\ref{dft2}):
\begin{equation}
\frac{\mathcal{P}_{t}(\dv N)}{\mathcal{P}_{t}(-\dv N)}=\left(\frac{\rho_{A}}{\rho_{B}}\right)^{\dv N}=\left(e^{\frac{\force_{N}}{k}}\right)^{\dv N}=e^{\dv S/k}.
\end{equation}

The calculation of the marginal probability distribution for the energy transport is slightly more involved.  We will invoke the independence of fluxes from each reservoir to the other one, as already alluded to when discussing the additivity property of the generating function. The total amount of energy $\dv U$ transfered during a certain time interval can be written as the difference of two independent contributions $\dv U_{A}$ and $\dv U_{B}$, being the energy coming from compartment $A$ and $B$ respectively:
\begin{equation}
\dv U=\dv U_{A}-\dv U_{B}.
\end{equation}
It follows that $\mathcal{P}_{t}(\dv U)$ can be expressed as a convolution integral:
\begin{equation}
\mathcal{P}_{t}(\dv U)=\int_{-\infty}^{+\infty}\mathcal{P}_t^{A}(E)\mathcal{P}_t^{B}(E-\dv U)\ud E,
\end{equation}
with $\mathcal{P}_t^{A}(\dv U)$ and $\mathcal{P}_t^{B}(\dv U)$ being the probability distributions to have an energy transfer $\dv U$ as a result of particles solely going from $A \rightarrow B$ and $B \rightarrow A$, respectively. From:
\begin{equation}
e^{-t\mu_{A}(\lambda_{U},0)}=\int_{-\infty}^{+\infty} e^{-\lambda_{U}\dv U }\mathcal{P}_t^{A}(\dv U) \ud \dv U ,
\end{equation}
the calculation of $\mathcal{P}_t^{A}(\dv U)$ involves an inverse Fourier transform, which is reproduced in Appendix \ref{appendixMD}. The final result reads:
\begin{widetext}
\begin{equation}\label{pau}
\mathcal{P}_t^{A}(\dv U)=e^{-t \sigma \rho_{A}\sqrt{\frac{kT_{A}}{2\pi m}}}\left\{
\delta(\dv U)+\theta(\dv U)\frac{t \sigma \rho_{A}}{\sqrt{2\pi m kT_{A}}}\frac{\dv U}{kT_{A}}e^{-\frac{\dv U}{kT_{A}}}\;_{0}F_{2}\left(\{\},\left\{\frac{3}{2},2\right\},\frac{t \sigma \rho_{A} \sqrt{kT_{A}}}{4\sqrt{2\pi m}} \left[\frac{\dv U}{kT_{A}}\right]^{2}\right)\right\}.
\end{equation}
\end{widetext}
The expression for $\mathcal{P}_t^{B}(\dv U)$ is obtained by replacing $A$ by $B$ in the above equation. When $T_{A}^{3/2}/\rho_A = T_{B}^{3/2}/\rho_B$, i.e., the thermodynamic force for the particle flow vanishes, $\mathcal{A}_{N}=0$, the resulting probability density $\mathcal{P}_t(\dv U)$ satisfies the fluctuation theorem Eq.~(\ref{dft2}). This follows directly by verifying from Eq.~(\ref{pau}) that 
$\mathcal{P}_t^{A}(\dv U)$ and $\mathcal{P}_t^{B}(\dv U)$ obey the relation ($\forall E$):
\begin{equation}
\frac{\mathcal{P}_t^{A}(E)\mathcal{P}_t^{B}(E-\dv U)}{\mathcal{P}_t^{A}(E-\dv U)\mathcal{P}_t^{B}(E)}=e^{\dv S/k}.
\end{equation}

\section{Large deviation function}
\begin{figure}[t]
\begin{center}
\includegraphics[width=0.4\textwidth]{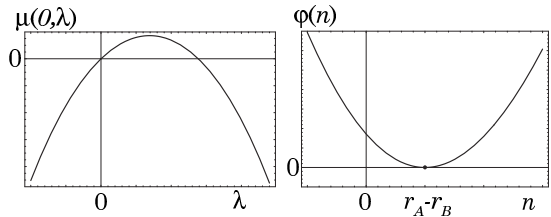}
\caption{Sketch of the generating function $\mu(0,\lambda)$ and its Legendre transform, the large deviation function $\varphi(n)$ for the case $r_{A}>r_{B}$.}
\label{ldf_N}
\end{center}
\end{figure}

The particle transport $\dv N$ is expected to grow proportional to time $t$. The asymptotic behaviour of $\mathcal{P}_{t}(\dv N)$ for large time $t \rightarrow \infty$ is  indeed described in terms of the variable $n \equiv {\dv N}/{t}$, namely:
\begin{equation}
\mathcal{P}_{t}(\dv N) \sim e^{-t\varphi(n)}.
\end{equation}
More precisely, the following limit:
\begin{equation}\label{ldf}
\varphi(n)=-\lim_{t \rightarrow \infty}\frac{1}{t}\ln \mathcal{P}_{t}(nt),
\end{equation}
is a convex function, independent of $t$, and  known as the large deviation function. This function can also be obtained as the Legendre transform of the generating function $\mu(0,\lambda)$:
\begin{equation}\label{lt}
\varphi(n)=\sup_{\lambda}\{\mu(0,\lambda)-\lambda n\}.
\end{equation}
Using following asymptotic expansion of the Bessel functions for large orders \cite{abramowitz}:
\begin{equation}
I_{\nu}(\nu z) \sim \frac{1}{\sqrt{2\pi \nu}}\frac{e^{\nu\left(\sqrt{1+z^{2}}+\ln \frac{z}{1+\sqrt{1+z^{2}}} \right) }}{(1+z^{2})^{1/4}} \mbox{ for $\nu \rightarrow \infty$},
\end{equation}
and substituting $\nu = nt$ and $z=2\sqrt{r_{A}r_{B}}/n$, we obtain from Eq.~(\ref{bessel}):
\begin{widetext}
\begin{equation}
\mathcal{P}_{t}(nt) \sim e^{-t(r_{A}+r_{B})} e^{nt\ln \sqrt{r_{A}/r_{B}}}\frac{1}{\sqrt{2\pi t}}
\frac{e^{t\left(\sqrt{4r_{A}r_{B}+n^{2}}+n\ln
\frac{2\sqrt{r_{A}r_{B}}} {n+\sqrt{4r_{A}r_{B}+n^{2}}}\right)}}
{(4r_{A}r_{B}+n^{2})^{1/4}}.
\end{equation}
\end{widetext}
This result leads to the following expression for $\varphi(n)$:
\begin{multline}
\varphi(n)=r_{A}+r_{B}-\sqrt{4r_{A}r_{B}+n^{2}} \\ -n\ln \left(\frac{\sqrt{4r_{A}r_{B}+n^{2}}-n}{2r_{B}}\right),
\end{multline}
which is of course the well-know large deviation function for a (biased) random walk. The same result is obtained using the Legendre transform of the generating function $\mu(0,\lambda)$. A sketch of both $\mu(0,\lambda)$ and $\varphi(n)$ is given in Fig.(\ref{ldf_N}). The function $\varphi(n)$ is positive everywhere, and has a single zero at the most probable value $n=\bar{n}=r_{A}-r_{B}$. 

\begin{figure}[t]
\begin{center}
\includegraphics[width=0.4\textwidth]{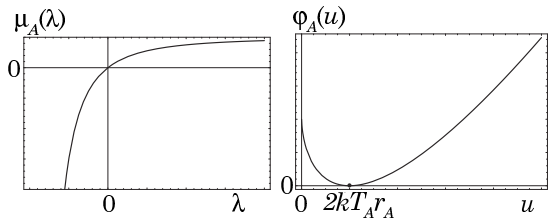}
\caption{Sketch of the generating function $\mu_{A}(\lambda)$ and the large deviation function $\varphi_{A}(u)$.}
\label{ldf_U}
\end{center}
\end{figure}

Similarly, the asymptotic properties for $t \rightarrow \infty$ of the energy transport are obtained by focusing on the scaled variable $u \equiv {\dv U}/{t}$. The large deviation function of $\mathcal{P}_t^{A}(\dv U)$,
\begin{equation}
\varphi_{A}(u)=-\lim_{t \rightarrow \infty}\frac{1}{t}\ln \mathcal{P}_{t}^A(ut),
\end{equation}
is again obtained either by Legendre transform of $\mu_{A}(\lambda,0)$, cf. Eq.~(\ref{mou}), or directly from Eq.~(\ref{pau}) in which case the asymptotic behaviour of $_{0}F_{2}(\{\},\{3/2,2\},z)$ is needed:
\begin{equation}
_{0}F_{2}(\{\},\{3/2,2\},z)\sim \frac{e^{3z^{1/3}}}{4\sqrt{3\pi}z^{5/6}} \mbox{ for $\vert z \vert \rightarrow \infty$}.
\end{equation}
The final result reads:
\begin{equation}
\varphi_{A}(u)=\frac{\sigma \rho_{A}\sqrt{kT_{A}}}{\sqrt{2\pi m}}+\frac{u}{kT_{A}}-\frac{3}{\sqrt{kT_{A}}}\left(\frac{\sigma \rho_{A}u^{2}}{4\sqrt{2\pi m}}\right)^{1/3},
\end{equation}
A sketch of the generating function $\mu_{A}(\lambda,0)$ and $\varphi_{A}(u)$ is given in Fig.(\ref{ldf_U}). $\varphi_{A}(u)$ has a unique zero at the most probable value $u=\bar{u}={2\sigma \rho_{A}(kT_{A})^\frac{3}{2}}/{\sqrt{2\pi m}}=2kT_{A}r_{A}$. This result is in agreement with the fact, cf. introduction, that particles carry on average $2kT_A$ energy.
Note that $\mathcal{P}_t^{A}(\dv U)$ decays exponentially for large $\dv U$, explaining the divergence of $\mu_A(\lambda,0)$ for  $\lambda \xrightarrow{>} -1/(kT_A)$, and the convergence of $\varphi_{A}(u)$ to a linear function for $u \rightarrow \infty$.

\section{Reciprocity relations}\label{reciprocity}
Averaging Eq.~(\ref{eq:entropy}) and taking the time derivative gives the average entropy production:
\begin{equation}
\frac{\ud}{\ud t}\langle \dv S \rangle=\ju \force_{U} +\jn \force_{N},
\end{equation}
where we define the macroscopic fluxes $\ju$ and $\jn$ corresponding to energy and particle transport, cfr. Eqs.~(\ref{cum1})-(\ref{cum2}): 
\begin{align}
\ju&=\frac{\ud}{\ud t}\langle \dv U \rangle=\frac{\sigma 2k^{3/2}}{\sqrt{2\pi m}}\left(\rho_{A}T_{A}^{3/2}-\rho_{B}T_{B}^{3/2} \right);
\label{fluxesEnergy} \\
\jn&=\frac{\ud}{\ud t}\langle \dv N \rangle=\frac{\sigma \sqrt{k}}{\sqrt{2\pi m}}\left(\rho_{A}T_{A}^{1/2}-\rho_{B}T_{B}^{1/2} \right).
\label{fluxesParticles}
\end{align}
In general, these fluxes are functions of the affinities $(\force_{U},\force_{N})$. Close to equilibrium, i.e. for small temperature and density differences $\dv T$ and $\dv \rho$, we can write:
\begin{eqnarray}
&&T_{A}=T-\dv T/2 \;\; \mbox{ ; }\;\; \rho_{A}=\rho-\dv \rho/2; \nonumber \\
&&T_{B}=T+\dv T/2 \;\; \mbox{ ; }\;\; \rho_{B}=\rho+\dv \rho/2,
\end{eqnarray}
so that the thermodynamic forces become:
\begin{equation}
\force_{U}\approx -\frac{\dv T}{T^{2}} \;\; \mbox{ and } \;\; \force_{N}\approx k\left(\frac{3}{2}\frac{\dv T}{T}-\frac{\dv \rho}{\rho}\right).
\end{equation}
A Taylor expansion of $\ju$ and $\jn$ leads to the following results:
\begin{widetext}
\begin{equation}
\ju=\frac{\sigma}{\sqrt{2\pi m}}2k^{1/2}\rho T^{3/2}\left(3kT\force_{U}+\force_{N}+\frac{9}{64}kT^{3}\force_{U}^{3}+\frac{3}{32}T^{2}\force_{U}^{2}\force_{N} +\frac{3}{2048}kT^{5}\force_{U}^{5}+\frac{3}{2048}T^{4}\force_{U}^{4}\force_{N}
+\ldots \right),
\end{equation}
and
\begin{equation}
\jn=\frac{\sigma}{\sqrt{2\pi m}}2k^{1/2}\rho T^{3/2}\left(\force_{U}+\frac{1}{2Tk}\force_{N}-\frac{1}{64}T^{2}\force_{U}^{3}-\frac{1}{64}\frac{T}{k}\force_{U}^{2}\force_{N}-\frac{1}{1024}T^{4}\force_{U}^{5}-\frac{5}{4096}\frac{T^{3}}{k}\force_{U}^{4}\force_{N}
+\ldots \right).
\end{equation}
\end{widetext}
From these expressions, it is clear that the Onsager symmetry \cite{onsager,prigogine,degroot} is satisfied:
\begin{equation}
\left.\frac{\partial \ju}{\partial \force_{N}} \right|_{0} =\left.\frac{\partial \jn}{\partial \force_{U}} \right|_{0}
\end{equation}
where the derivatives are evaluated at equilibrium $(\force_{U},\force_{N})=(0,0)$. This relation is in fact also a direct consequence of the symmetry relation of $\mu(\lambda_{U},\lambda_{N})$, cfr. Eq.~(\ref{DFT}) \cite{lebowitz,gallavotti2,gaspard1,gaspard2}. To show this explicitly, it is convenient to include the dependence of $\mu$ on the  thermodynamic forces, i.e. we write $\mu(\lambda,\force)$ (for ease of notation, we use $\lambda=\{\lambda_{U},\lambda_{N}\}$ and $\force=\{\force_{U},\force_{N}\}$). We then find:
\begin{align}
\left.\frac{\partial \ju}{\partial \force_{N}} \right|_{0}&=\left.\frac{\partial^{2}\mu(\lambda,\force)}{\partial \force_{N} \partial \lambda_{U}}\right|_{0}=\left.\frac{\partial^{2}\mu(\force/k-\lambda,\force)}{\partial \lambda_{U}\partial \force_{N}}\right|_{0}\\
&=-\frac{1}{k}\left.\frac{\partial^{2}\mu(\lambda,0)}{\partial \lambda_{U}\partial \lambda_{N}}\right|_{0}
+\left.\frac{\partial^{2}\mu(-\lambda,\force)}{\partial \lambda_{U}\partial \force_{N}}\right|_{0}.
\end{align}
The last term is again $-\left.\partial \ju / \partial \force_{N} \right|_{0}$, so that:
\begin{equation}
\left.\frac{\partial \ju}{\partial \force_{N}} \right|_{0}=
-\frac{1}{2k}\left.\frac{\partial^{2}\mu(\lambda,0)}{\partial \lambda_{U}\partial \lambda_{N}}\right|_{0},
\end{equation}
which leads to the Onsager symmetry relation.

\section{2D results}
All the above results have been derived for three spatial dimensions. For comparison with computer simulations, we reproduce the corresponding results for two spatial dimensions.
We note the following  two main modifications:
\begin{itemize}
\item[i)] The thermodynamic force corresponding to the particle flux has the following form:
\begin{equation}
\force_{N}=k\log\left[\frac{\rho_{A}T_{B}}{\rho_{B}T_{A}}\right].
\end{equation}
\item[ii)] The transition rates $T_{A \rightarrow B}(E)$ and $T_{B \rightarrow A}(E)$ are to be calculated using the two-dimensional Maxwellian velocity distribution:
\begin{equation}
\phi_{i}(\vec{v})=\frac{m}{2\pi kT_{i}}e^{-\frac{mv^{2}}{2kT_{i}}},
\end{equation}
\end{itemize}
namely:
\begin{widetext}
\begin{equation}
T_{A \rightarrow B}(E)=\int_{v=0}^{\infty}\int_{\varphi=0}^{\pi}v \ud v \ud \varphi\; \rho_{A}\phi_{A}(\vec{v})\;\sigma \;v\sin \varphi \;
\delta (E-\frac{1}{2}mv^{2})=\frac{\sqrt{2}\sigma \rho_{A}}{\pi \sqrt{m k T_{A}}}\left(\frac{E}{kT_{A}}\right)^{\frac{1}{2}}e^{-\frac{E}{kT_{A}}} ;
\end{equation}
and
\begin{equation}
T_{B \rightarrow A}(E)=\int_{v=0}^{\infty}\int_{\varphi=\pi}^{2\pi}v \ud v \ud \varphi\; \rho_{B}\phi_{B}(\vec{v})\;\sigma \;(-v\sin \varphi) \;
\delta (E-\frac{1}{2}mv^{2})=\frac{\sqrt{2}\sigma \rho_{B}}{\pi \sqrt{m k T_{B}}}\left(\frac{E}{kT_{B}}\right)^{\frac{1}{2}}e^{-\frac{E}{kT_{B}}}.
\end{equation}
\end{widetext}
They display a  prefactor proportional to $\sqrt{E}$, hence transitions with $E < kT$ gain more weigth as compared to the $3$D situation. The calculation of the generating function proceeds in exactly the same way. The final result differs from the $3$D case only by the $3/2$ power law exponent: 
\begin{multline}
\mu(\lambda_{U},\lambda_{N})=\frac{\sigma \sqrt{k}}{\sqrt{2\pi m}}\left(\rho_{A}\sqrt{T_{A}}\left[1-\frac{e^{-\lambda_{N}}}{(1+kT_{A}\lambda_{U})^{3/2}}\right]\right. \\
\left.+\rho_{B}\sqrt{T_{B}}\left[1-\frac{e^{\lambda_{N}}}{(1-kT_{B}\lambda_{U})^{3/2}}\right] \right).
\end{multline}
The fluctuation relation Eq.~(\ref{DFT}) is again verified. The first and second order cumulants read:
\begin{align}
\kappa_{10}&=\langle \dv U \rangle= \frac{3}{2}\frac{t \sigma k^{{3/2}}}{\sqrt{2\pi m}}\left(\rho_{A}T_{A}^{3/2}-\rho_{B}T_{B}^{3/2} \right);\\
\kappa_{01}&=\langle \dv N \rangle= \frac{t \sigma \sqrt{k}}{\sqrt{2\pi m}}\left(\rho_{A}T_{A}^{1/2}-\rho_{B}T_{B}^{1/2} \right);\\
\kappa_{20}&=\langle \delta \dv U^{2}\rangle= \frac{15}{4}\frac{t \sigma k^{5/2}}{\sqrt{2\pi m}}\left(\rho_{A}T_{A}^{5/2}+\rho_{B}T_{B}^{5/2} \right);\\
\kappa_{11}&= \frac{3}{2}\frac{t \sigma k^{3/2}}{\sqrt{2\pi m}}\left(\rho_{A}T_{A}^{3/2}+\rho_{B}T_{B}^{3/2} \right);\\
\kappa_{02}&=\langle \delta \dv N^{2} \rangle= \frac{t \sigma \sqrt{k}}{\sqrt{2\pi m}}\left(\rho_{A}T_{A}^{1/2}+\rho_{B}T_{B}^{1/2} \right).
\end{align}

\section{2D Molecular Dynamics}
\begin{figure}
\begin{center}
\includegraphics[width=0.40\textwidth]{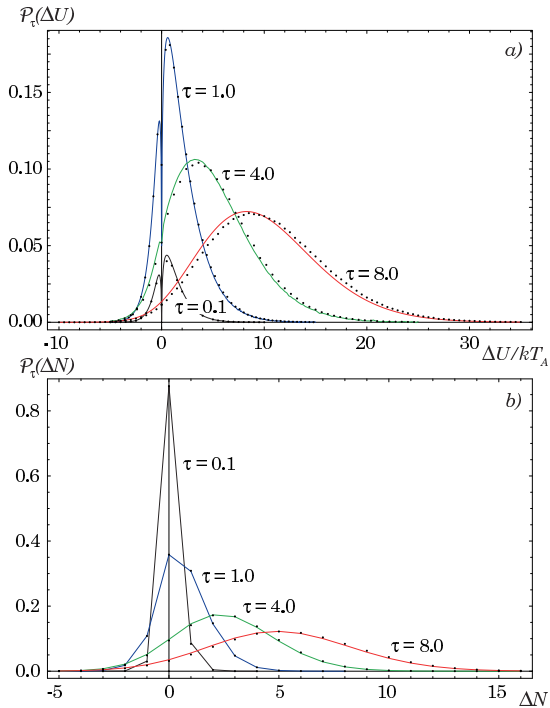}
\caption{(Color online) Comparison between theory and simulation with $\rho_{A}/\rho_{B}=2.0$ and $T_{A}/T_{B}=2.0$ ($N_{A}=2N_{B}=2000$ and $T_{A}=2T_{B}=1$) for different time intervals. a) Plot of the marginal distribution $\mathcal{P}_{\tau}(\dv U)$ (a Dirac Delta contribution at the origin is omitted). b) Plot of the marginal distribution $\mathcal{P}_{\tau}(\dv N)$. Dots: molecular dynamics simulations; full line: theoretical results (in b) they serve as a guide to the eye).}
\label{figPdfu2D}
\end{center}
\end{figure}

\begin{figure}
\begin{center}
\includegraphics[width=0.40\textwidth]{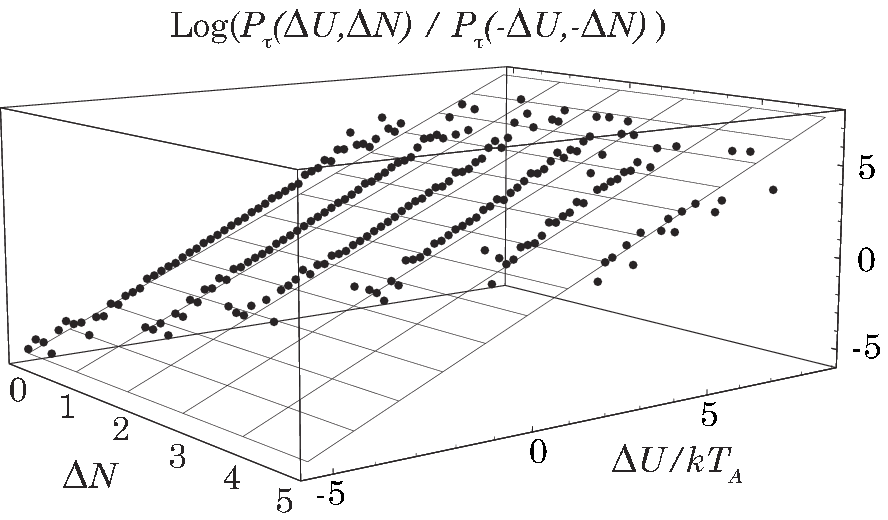}
\caption{2D Plot of $\log (P_{\tau}(\dv U,\dv N)/P_{\tau}(-\dv U,-\dv N))$ for $\rho_{A}/\rho_{B}=2.0$, $T_{A}/T_{B}=2.0$ ($N_{A}=2N_{B}=2000$ and $T_{A}=2T_{B}=1$) and $\tau=1.0$.}
\label{figCrooks2D}
\end{center}
\end{figure}

\begin{figure}
\begin{center}
\includegraphics[width=0.40\textwidth]{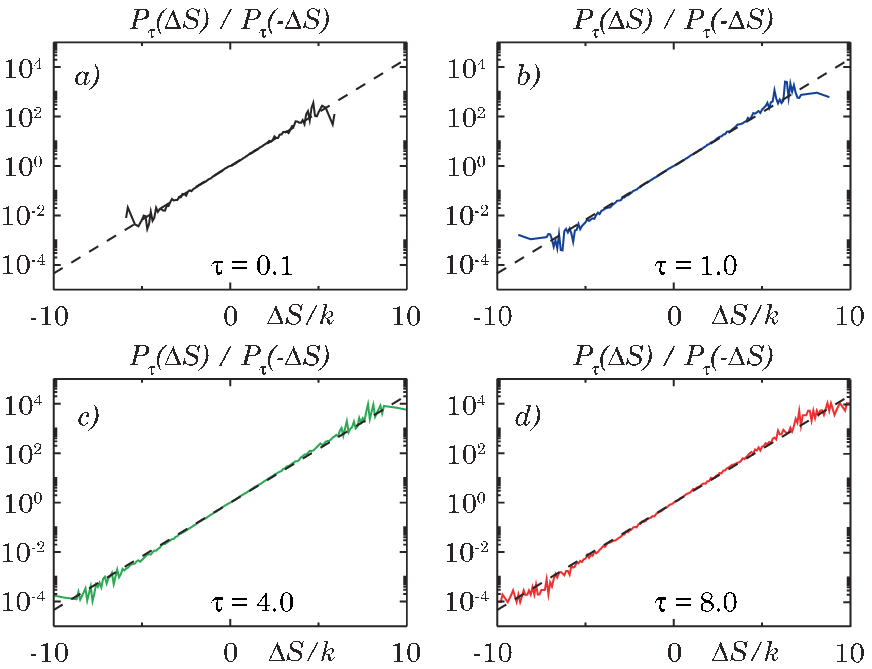}
\caption{(Color online) Plot of $P_{\tau}(\dv S)/P_{\tau}(-\dv S)$ for $\rho_{A}/\rho_{B}=2.0$ and $T_{A}/T_{B}=2.0$ ($N_{A}=2N_{B}=2000$ and $T_{A}=2T_{B}=1$) for different time intervals: a) $\tau=0.1$, b) $\tau=1.0$, c) $\tau=4.0$, d) $\tau=8.0$. The dashed line shows the theoretical result $\exp(-\Delta S/k)$.}
\label{figCrooksDS}
\end{center}
\end{figure}

\begin{figure}
\begin{center}
\includegraphics[width=0.40\textwidth]{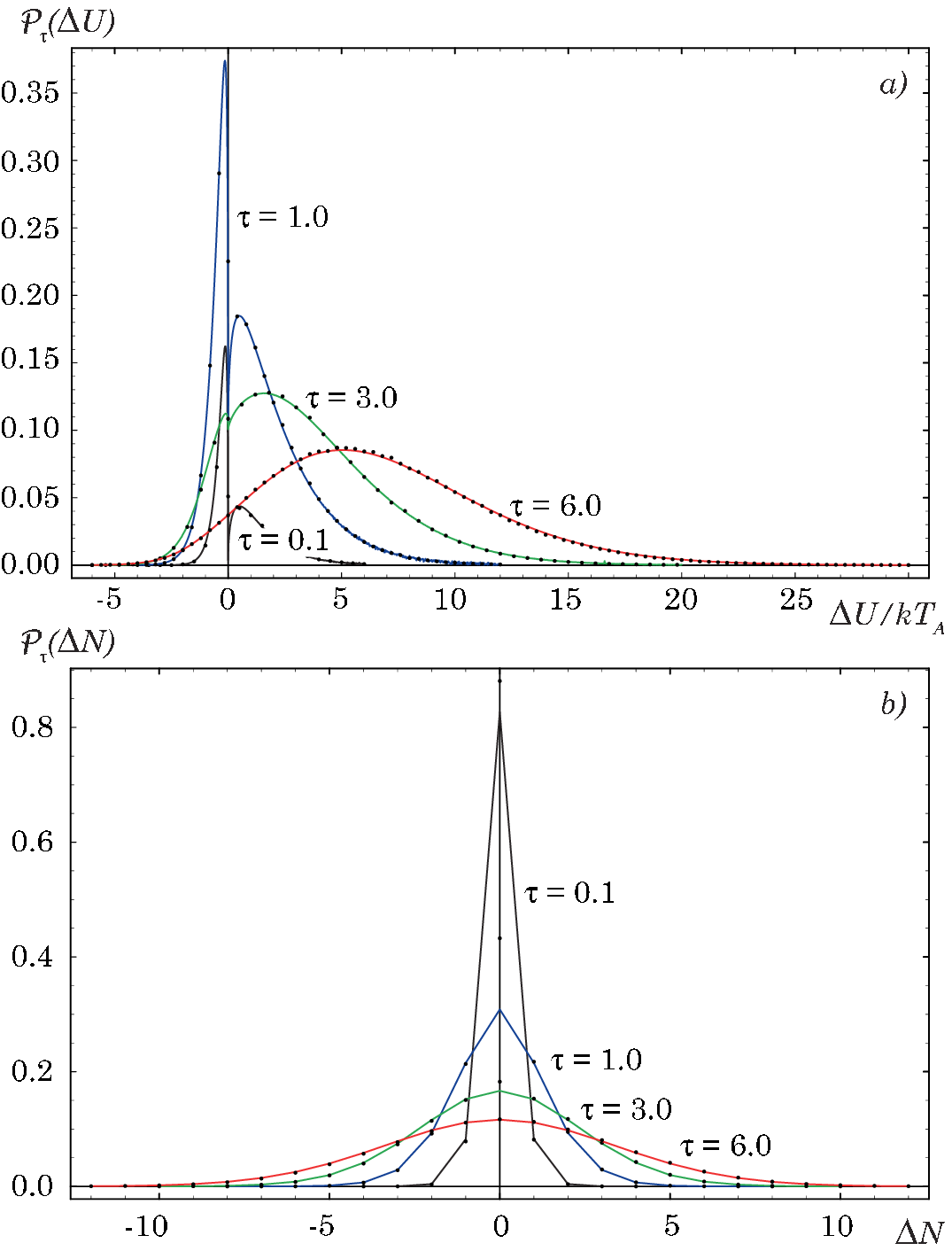}
\caption{(Color online) Comparison between theory and simulation with $\rho_{A}/\rho_{B}=0.5$ and $T_{A}/T_{B}=4.0$ ($2N_{A}=N_{B}=2000$ and $T_{A}=4T_{B}=1$) for different time intervals. a) Plot of the marginal distribution $\mathcal{P}_{\tau}(\dv U)$ (a Dirac Delta contribution at the origin is ommitted). b) Plot of the marginal distribution $\mathcal{P}_{\tau}(\dv N)$. Dots: molecular dynamics simulations; full line: theoretical results (in b) they serve as a guide to the eye). Note that the average particle flux is zero ($\rho_{A}\sqrt{T_{A}}=\rho_{B}\sqrt{T_{B}}$).}
\label{figPdfu2Dn}
\end{center}
\end{figure}

\begin{figure}
\begin{center}
\includegraphics[width=0.40\textwidth]{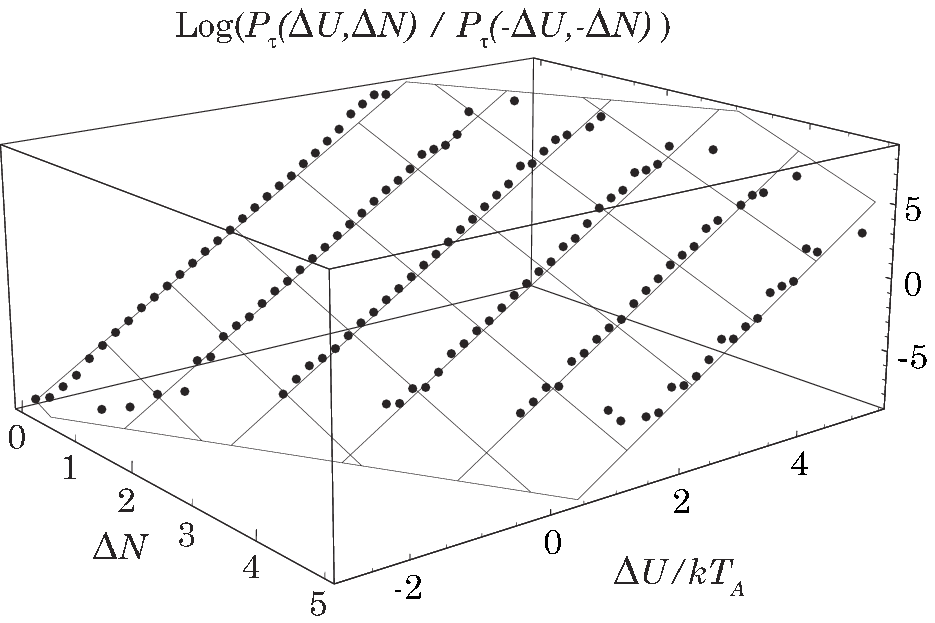}
\caption{2D Plot of $\log (P_{\tau}(\dv U,\dv N)/P_{\tau}(-\dv U,-\dv N))$ for $\rho_{A}/\rho_{B}=0.5$, $T_{A}/T_{B}=4.0$ ($2N_{A}=N_{B}=2000$ and $T_{A}=4T_{B}=1$) and $\tau=1.0$.}
\label{figCrooks2Dn}
\end{center}
\end{figure}

\begin{figure}
\begin{center}
\includegraphics[width=0.40\textwidth]{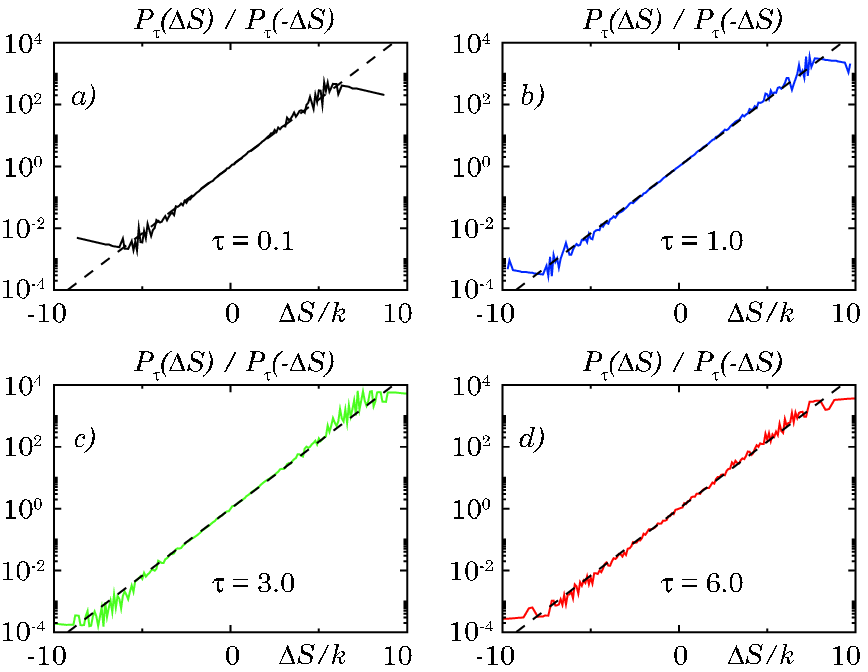}
\caption{(Color online) Plot of $P_{\tau}(\dv S)/P_{\tau}(-\dv S)$ for $\rho_{A}/\rho_{B}=0.5$ and $T_{A}/T_{B}=4.0$ ($2N_{A}=N_{B}=2000$ and $T_{A}=4T_{B}=1$) for different time intervals: a) $\tau=0.1$, b) $\tau=1.0$, c) $\tau=3.0$, d) $\tau=6.0$. The dashed line shows the theoretical result $\exp(-\Delta S/k)$.}
\label{figCrooksDSn}
\end{center}
\end{figure}

Extensive molecular dynamics simulations were caried out for the 2D system. Two equally large reservoirs of total size $5000\times 200$ (reflecting boundary conditions) and connected through a hole of diameter $\sigma=5$, contain dilute gases of hard disks of diameter d = 1 and mass m = 1. Initially, the hole between the reservoirs is closed and $N_{A}$ ($N_{B}$) disks are randomly placed in reservoir $A$ ($B$). Their velocities are randomly sampled from the Maxwellian velocity distribution at the corresponding temperature and slightly rescaled in order to make the center-of-mass velocity zero in both reservoirs, and the total energy equal to $N_{A}kT_{A}$ and $N_{B}kT_{B}$ in reservoir $A$ and $B$ respectively. Then, the system is allowed to relax for a certain period, after which the hole is opened and $\Delta U$ and $\Delta N$ are measured. Averages are taken over $1\;000\;000$ realizations. The duration time $\tau$ of the simulation is measured in units of the average time between two particles going from $A$ to $B$, and is related to $t$ via: 
\begin{equation}
\tau=\frac{t \sigma \rho_{A}\sqrt{kT_{A}}}{\sqrt{2\pi m}}.
\end{equation}
A comparison with the above given theoretical results for an ideal gas is shown in Figs.~\ref{figPdfu2D},\ref{figCrooks2D},\ref{figCrooksDS},\ref{figPdfu2Dn},\ref{figCrooksDSn} and \ref{figCrooks2Dn} and Tables~\ref{tbl:moments} and \ref{tbl:momentsn}. Agreement is extremely good. In particular, the fluctuation theorem is verified, cf. Eq.~\ref{ft}. In Table~\ref{tbl:moments} and \ref{tbl:momentsn}, we have also included the numerically obtained value of the average $\langle \exp{-\Delta S/k}\rangle$. Note the deviations for increasing $\tau$ from the exact theoretical value $1$ due to unsufficient sampling of the negative tails of $P_{t}(\Delta S)$.
 
\begingroup
\squeezetable
\begin{table*}
\caption{Comparison between molecular dynamics simulation and theory ($\rho_{A}/\rho_{B}=2.0$ and $T_{A}/T_{B}=2.0$).} 
\label{tbl:moments}
\begin{ruledtabular}
\begin{tabular}{dddddddddddd}
\multicolumn{1}{c}{$\tau$} &
\multicolumn{2}{c}{$\kappa_{10}$}& \multicolumn{2}{c}{$\kappa_{01}$} &
\multicolumn{2}{c}{$\kappa_{20}$} & \multicolumn{2}{c}{$\kappa_{11}$}  &
\multicolumn{2}{c}{$\kappa_{02}$} & \multicolumn{1}{c}{$\langle e^{-\Delta S/k} \rangle$} \\
&\multicolumn{1}{c}{sim.}
&\multicolumn{1}{c}{theory}&\multicolumn{1}{c}{sim.}
&\multicolumn{1}{c}{theory}&\multicolumn{1}{c}{sim.}
&\multicolumn{1}{c}{theory}&\multicolumn{1}{c}{sim.}
&\multicolumn{1}{c}{theory}&\multicolumn{1}{c}{sim.}
&\multicolumn{1}{c}{theory}&\multicolumn{1}{c}{sim.}\\
\colrule
0.1 & 0.122&0.124&0.062&0.065& 0.407& 0.408& 0.172& 0.177& 0.127& 0.135 &0.997\\
1.0 & 1.280&1.235&0.657&0.647& 4.214& 4.082& 1.802& 1.765& 1.358& 1.354 &0.955\\
4.0 & 5.178&4.939&2.667&2.586&16.904&16.326& 7.247& 7.061& 5.456& 5.414 &0.773\\
8.0 &10.393&9.879&5.378&5.172&33.619&32.652&14.463&14.121&10.893&10.828 &0.707\\
\end{tabular}
\end{ruledtabular}
\end{table*}
\endgroup

\begingroup
\squeezetable
\begin{table*}
\caption{Comparison between molecular dynamics simulation and theory ($\rho_{A}/\rho_{B}=0.5$ and $T_{A}/T_{B}=4.0$).} 
\label{tbl:momentsn}
\begin{ruledtabular}
\begin{tabular}{dddddddddddd}
\multicolumn{1}{c}{$\tau$} &
\multicolumn{2}{c}{$\kappa_{10}$}& \multicolumn{2}{c}{$\kappa_{01}$} &
\multicolumn{2}{c}{$\kappa_{20}$} & \multicolumn{2}{c}{$\kappa_{11}$}  &
\multicolumn{2}{c}{$\kappa_{02}$} & \multicolumn{1}{c}{$\langle e^{-\Delta S/k} \rangle$} \\
&\multicolumn{1}{c}{sim.}
&\multicolumn{1}{c}{theory}&\multicolumn{1}{c}{sim.}
&\multicolumn{1}{c}{theory}&\multicolumn{1}{c}{sim.}
&\multicolumn{1}{c}{theory}&\multicolumn{1}{c}{sim.}
&\multicolumn{1}{c}{theory}&\multicolumn{1}{c}{sim.}
&\multicolumn{1}{c}{theory}&\multicolumn{1}{c}{sim.}\\
\colrule
0.1&0.113&0.113&0.004&0.0& 0.191& 0.398& 0.184& 0.188& 0.395& 0.2 &0.988\\
1.0&1.138&1.125&0.015&0.0& 3.984& 3.984& 1.896& 1.875& 1.998& 2.0 &0.910\\
3.0&3.375&3.375&0.040&0.0&11.755&11.953& 5.646& 5.625& 5.981& 6.0 &0.515\\
6.0&6.679&6.750&0.083&0.0&23.145&23.906&11.226&11.250&11.920&12.0 &0.214\\
\end{tabular}
\end{ruledtabular}
\end{table*}
\endgroup

\section{Discussion}
The results presented here, the fluctuation theorem for the effusion of an ideal gas, have a certain academic and didactic appeal. The underlying physics is easy to grasp, while the exact and detailed analytic solution completes the understanding. The results are not restricted to the regime of linear response since the gradients in density or temperature between the reservoirs can be arbitrarily
large. Yet, a few words of caution are in place. First, the present example is only a limited illustration of the fluctuation theorem. By considering effusion through a small hole, the system is at all times in local equilibrium. In this case, the fluctuation theorem can in fact be derived using equilibrium concepts only. However the fluctuation theorem remains valid away from local equilibrium, for example when the opening between the reservoirs is much larger than the mean free path \cite{newpaper}. Second, either due to the smallness of the hole or due to the ideality of the gas, we did not need to discuss the energy required to open and close the hole connecting the reservoirs. The size and impact of this contribution on the fluctuation theorem will obviously depend on the system under consideration \cite{farago1,vanzon,wojcik,sancho,visco}. Finally,  fluctuation and work theorems can also be formulated for time-dependent situations, rather than the stationary state that was discussed here. The trademark of these theorems is that one has to distinguish probability distributions (for entropy or work) for direct and time reversed schedules. These probabilities are typically not identical and in fact can describe quite different physical situations of the system, see for example \cite{cleurenPRL2006}.

\appendix
\section{Transition rates}\label{transition}
\begin{figure}
\begin{center}
\includegraphics[width=0.3\textwidth]{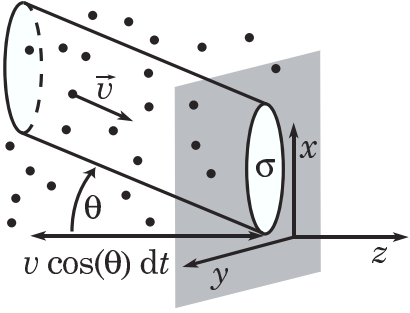}
\caption{Sketch of the gas near the hole. A particle with velocity $\vec{v}$ in must be located inside the tilted cilinder with volume $\sigma v\cos \theta \ud t$, in order to reach the hole during the time interval $\ud t $.}
\label{figappendixA}
\end{center}
\end{figure}

The calculation of the transition rates as given in Eq.~(\ref{eq:tr}), is similar to that of the pressure in textbooks on kinetic theory. We focus here on $T_{A \rightarrow B}(E)$, since the result for $T_{B \rightarrow A}(E)$ is obtained from it by replacing $A\rightarrow B$. In the following, we write the velocity in spherical coordinates, i.e. $\vec{v}=(v\sin \theta \cos \varphi,v\sin \theta \sin \varphi,v\cos \theta)$, with the $z$-axis perpendicular to the common wall between the two reservoirs, and pointing from $A$ to $B$, see Fig. \ref{figappendixA}.\newline $T_{A \rightarrow B}(E)\ud E \ud t$ is the probability to observe a particle with kinetic energy in the range $]E,E+\ud E[$, crossing the hole from $A$ to $B$ during a time interval $\ud t$. Two requirements must be fullfilled in order to observe this crossing. First, the kinetic energy of the particle must be in the specified energy range, which means that its velocity is in the range $]v,v+\ud v[$ with:
\begin{equation}
v=\sqrt{2E/m}\;\;\;\; \mbox{and}\;\;\;\;\ud v=\ud E/\sqrt{2Em}.
\end{equation}
Second, the particle must be able to reach the hole during the time interval $\ud t$. A particle with velocity $v\cos \theta$ in the $z$-direction must therefore be located inside the volume $\sigma v\cos \theta \ud t$ (cfr. Fig. \ref{figappendixA}). Since the $z$-component of the velocity must be positive if the particle is to move from $A$ to $B$; $\theta$ has to be between $0$ and $\pi/2$. \newline Adding all contributions from the different directions $\theta$ and $\varphi$ and noting that the velocities are distributed according to the Maxwellian velocity distribution $\phi_{A}(\vec{v})$, leads to the following result:
\begin{widetext}
\begin{eqnarray}
T_{A \rightarrow B}(E)\ud E \ud t &=&\int_{\theta=0}^{\pi /2}\int_{\varphi=0}^{2\pi}v^2 \sin \theta \ud v\ud \theta \ud \varphi\; \rho_{A}\phi_{A}(\vec{v})\;\sigma v\cos \theta \ud t \nonumber \\
&=&\int_{\theta=0}^{\pi /2}\int_{\varphi=0}^{2\pi}\frac{2E}{m} \sin \theta \frac{\ud E}{\sqrt{2Em}} \ud \theta \ud \varphi\; \rho_{A} \left(\frac{m}{2\pi kT_{A}}\right)^{\frac{3}{2}}e^{-\frac{E}{kT_{A}}}\;\sigma \sqrt{\frac{2E}{m}} \cos \theta  \ud t \nonumber \\
&=&\frac{\sigma \rho_{A}}{\sqrt{2\pi m k T_{A}}}\frac{E}{kT_{A}}e^{-\frac{E}{kT_{A}}}\ud E \ud t. 
\end{eqnarray}
\end{widetext}

\section{Calculation of the marginal distribution}\label{appendixMD}
Our starting point in the calculation of the marginal probability distribution $\mathcal{P}_{t}^{A}(\dv U)$ is the generating function $\mu_{A}(\lambda,0)$, cf. Eq.~(\ref{eq:mua}). For ease of notation, we use the following dimensionless quantities: 
\begin{equation}
X=\frac{\dv U}{kT_{A}};\tau=\frac{t \sigma \rho_{A}\sqrt{kT_{A}}}{\sqrt{2\pi m}}; \rho=\frac{\rho_{A}}{\rho_{B}};\chi=\frac{T_{A}}{T_{B}},
\end{equation}
which are respectively the energy measured in terms of the thermal energy in part $A$, the average time between two particles going from $A$ to $B$, the ratio of the densities and the ratio of the temperatures. In these new variables, the generating function becomes:
\begin{equation}
\mu_{A}(\lambda,0)=1-\frac{1}{(1+\lambda)^{2}}.
\end{equation}
Taking into account the Jacobian of the transformation form $\dv U \rightarrow X$, and defining $\mathcal{P}_{t}^{A}(\dv U)\ud \dv U=f_{\tau}(X)\ud X$ we obtain:
\begin{equation}
e^{-\tau \left(1-\frac{1}{(1+\lambda)^{2}}\right)}=\int_{\infty}^{\infty}e^{-\lambda X}f_{\tau}(X)\ud X.
\end{equation}
By inverse Fourier transform (setting $\lambda = iq)$, we find:
\begin{equation}
f_{\tau}(X)=\frac{1}{2\pi}\int_{\infty}^{\infty}e^{-\tau \left(1-\frac{1}{(1+iq)^{2}}\right)}e^{iqX}\ud q.
\end{equation}
Under the integral, we add and substract the function $e^{-\tau}e^{iqX}$. The purpose for doing is twofold. First, it allows to single out the Dirac delta at the origin:
\begin{equation}
f_{\tau}(X)=e^{-\tau}\delta(X)+\frac{e^{-\tau}}{2\pi}\int_{\infty}^{\infty}\left(e^{\frac{\tau}{(1+iq)^{2}}}-1\right)e^{iqX}\ud q.
\end{equation}
Second, the remaining integral can be converted to a contour integral using Jordan's Lemma by noting that the complex function $\exp{\frac{\tau}{(1+iz)^{2}}}-1$ goes to zero for large values of $\vert z \vert$. Then, for $X>0$ $(X<0)$, we can close the contour in the upper (lower) half plane. Since the integrand has only one (essential) singularity at $z=i$, i.e. in the upper half plane, and using the residue theorem it follows that $f_{\tau}(X<0)=0$, a result that was expected since a particle going from $A$ to $B$ always takes a positive amount of energy with it (its kinetic energy). Using the Heaviside function $\theta(x)$ we conclude:
\begin{align}
f_{\tau}(X)&=e^{-\tau}\delta(X)+\theta(X)\frac{e^{-\tau}}{2\pi}\oint
\left(e^{\frac{\tau}{(1+iz)^{2}}}-1\right)e^{izX}\ud z \nonumber \\
&=e^{-\tau}\delta(X)+\theta(X)e^{-\tau}i \mbox{Res}(i).
\end{align}
The residue of the integrandum is easily obtained by making a Laurent expansion around $z=i$:
\begin{multline}
\left(e^{\frac{\tau}{(1+iz)^{2}}}-1\right)e^{izX}
\\ =e^{-X}\sum_{j=-\infty}^{\infty}\left(\sum_{n=1}^{\infty}\frac{(-\tau)^{n}}{n!}\frac{(iX)^{(2n+j)}}{(2n+j)!}\right)(z-i)^{j} .
\end{multline}
And so:
\begin{equation}
f_{\tau}(X)=e^{-\tau}\delta(X) +\theta(X)e^{-(\tau+X)}\sum_{n=1}^{\infty}\frac{\tau^{n}}{n!}\frac{X^{(2n-1)}}{(2n-1)!}.
\end{equation}
Finally, this last summation can be done explicitly in terms of the generalized hypergeometric function $_{0}F_{2}$\footnote{In general, these functions are defined as $_{p}F_{q}(\{a_1,\ldots ,a_p\},\{b_1,\ldots ,b_q\},x)=\sum_{k=0}^{\infty}\frac{(a_1)_k\ldots (a_p)_k}{(b_1)_k\ldots (b_q)_k}\frac{x^{k}}{k!}$ where $(x)_n=\Gamma(x+n)/\Gamma(x)$ denotes the Pochhammer symbol}:
\begin{multline}
f_{\tau}(X)=e^{-\tau}\delta(X)
\\+\theta(X)e^{-(\tau+X)}\tau X\;_{0}F_{2}\left(\{\},\{\frac{3}{2},2\},\frac{\tau X^{2}}{4}\right).
\end{multline}
The final expression for $\mathcal{P}_{t}^{A}(\dv U)$ is given in Eq.~(\ref{pau}).

\bibliography{cleuren}

\end{document}